\documentclass{article}
\usepackage{graphicx}
\usepackage{amssymb}

\def\be{\begin{eqnarray}}
\def\ee{\end{eqnarray}}

\def\p{\partial}

\hoffset= - 1.10in         

\begin{document}

\hfill ITEP/TH-32/05

\hfill INR/TH-2005-19

\bigskip

\centerline{\Large{A Solvable Sector of AdS Theory.
}}

\bigskip
\centerline{D.Krotov$^{a}$ and A.Morozov$^{b}$}

\begin{center}
$^a${\small{\em
Institute for Nuclear Research of the Russian Academy of Sciences, }}\\
{\small{\em
60th October Anniversary prospect 7a, Moscow, 117312, Russia
}}\\
$^{a,b}${\small{\em
Institute of Theoretical and Experimental Physics, }}\\
{\small{\em
B.Cheremushkinskaya 25, Moscow, 117259, Russia
}}\\
$^a${\small{\em
Moscow State University, Department of Physics,}}\\
{\small{\em
Vorobjevy Gory, Moscow, 119899, Russia
}}
\end{center}
\bigskip

\centerline{ABSTRACT}

\bigskip

Field theory in space-time with boundary has an interesting
sub-sector, where propagator is {\it difference} of those with
Neumann and Dirichlet boundary conditions. Such boundary-induced
theory in the bulk is essentially holomorphic and is exactly
solvable in the sense that  all orders of perturbation theory
can be summed up explicitly into effective non-local theory at the
boundary. This provides a non-trivial realization of holography principle.
In the particular example of scalar fields of dimensions
$\Delta_\pm = (d\pm 1)/2$ in $AdS_{d+1}$ the corresponding effective
conformal theory has propagators $|\vec p\ |^{-1}$ and vertices
$\Big(|\vec p_1| + \ldots + |\vec p_n|\Big)^{-s_n}$ of valence $n$
in momentum representation, with $s_n = (n-2)\Delta_- - 1$. This
extraordinary simplicity of certain amplitudes in AdS seems
inspiring and can be helpful for analyzing corollaries of
open-closed string duality for particular field-theory sub-sectors
of string theory.

\bigskip

\bigskip

\section{Introduction}

Quantum field theory in the space-time with boundaries
has additional structures as compared to the theory when no
boundaries are present \cite{CS}-\cite{didi}.
One of the most important questions in this context is
relation between the theories in the bulk and at the
boundary, where the latter one describes the eigen-states
(wave-functions) of the evolution operators of the former.
This relation is believed to be especially important
\cite{holb,hole}
in topological and gravitational bulk theories with trivial
Hamiltonians. Further, it is expected to be promoted to
an even more fundamental open-closed string duality
\cite{Gopak}.

In this paper we consider a small subject
in the story of bulk-to-boundary correspondence, which
-- to the best of our knowledge -- did not attract enough
attention in the literature. It concerns a possibility to
form a nearly topological theory in the bulk, by
substituting the bulk propagators by {\it differences} of
those with Neumann and Dirichlet boundary conditions at
the boundary. Such subtraction eliminates the physical pole
from the propagator and -- if the boundary consists of the
stable points of discrete $Z_2$ transformation --
substitutes it by "unphysical" pole at $Z_2$-image of the
physical one. This situation is modeled by the theory
of complex scalar field $\chi$ in the upper semi-space
$R^{d+1}_+$ and the boundary $R^d$ located at $z_0 = 0$,
with partition function
$$
{\cal Z}\{J\} = \int
D\chi(z) D\bar\chi(z)
\exp \left\{-\frac{1}{g^2} \int_{R^{d+1}_+}
\left(\p_\mu \chi \p_\mu \bar\chi +
\sum_{n} \frac{t_n}{n!}\ z_0^{s_n-1}
\chi^n\right) d^{d+1}z +
\right. \vspace{-0.5cm}
$$\be
\left. \phantom{5^{5^{5^{5^{5^5}}}}} +
\frac{1}{g^2} \oint_{R^d} 
\Big(J_N(\vec x){\rm Re}\ \chi(\vec x) +
J_D(\vec x){\rm Im} \ \partial_0\chi(\vec
x)\Big)d\vec x \right\}
\label{bulk}
\ee
where functional integral is over
the fields with mixed (Neumann-Dirichlet) boundary
conditions
\be \left.\begin{array}{c}
{\rm Re}\ \partial_0\chi(\vec x, z_0) = 0\\
{\rm Im}\ \chi(\vec x, z_0) = 0
\end{array}\right|_{z_0 = 0}
\label{bc} \ee and $\vec x$ are coordinates on the $d$-dimensional
boundary (while coordinates in the bulk are $z_0,\vec z$). The
summation over $\mu$ is with respect to the flat Euclidean metric.
Be there no boundary, the theory would be trivial, since
propagator converts $\chi$ into $\bar\chi$, while interaction
terms are pure holomorphic: contain only $\chi(z)$, not
$\bar\chi(z)$. Because of the boundary conditions the cancellation
between the contributions of propagating ${\rm Re}\ \chi(z) =
\frac{1}{2}(\chi + \bar\chi)(z)$ and ${\rm Im}\ \chi(z) =
\frac{1}{2i}(\chi - \bar\chi)(z)$ is not complete and partition
function appears to be non-trivial. Still, it is particularly
simple, and -- as we demonstrate below -- can be straightforwardly
rewritten as an effective field theory of a single real field
$\varphi(\vec x)$ at the boundary, \be {\cal Z}\{J\} = \int
D\varphi(\vec x) \exp \left\{-\frac{1}{g^2} \oint_{R^d}
\left(\frac{1}{2}\varphi \sqrt{\Box} \varphi + \sum_{n}
\frac{t_n}{n!} \int_0^\infty
\left(e^{-\alpha\sqrt{\Box}}\varphi\right)^n \alpha^{s_n-1}
d\alpha + \Big(J_N + J_D\sqrt{\Box}\Big)\varphi \right) d\vec x
\right\} \label{bound} \ee Here $\Box = -\partial_{\vec
x}\partial_{\vec x}$ is {\it minus} Laplacian at the boundary and
$\alpha$ is {\it not} a field, just a single auxiliary
integrational variable.

The field $\bar\chi$ enters linearly in the action
 (\ref{bulk}) and can work as Lagrange multiplier,
providing a functional delta-function
$\delta\Big((\partial_0^2-\Box)\chi(z) +
J(\vec x)\delta(z_0)\Big)$. However, in the presence
of boundary this condition does not fix $\chi(z)$
unambiguously in the bulk: a {\it functional} freedom
remains in the zero-modes of the Laplace operator
in $R^{d+1}_+$ and it is not eliminated by our
mixed boundary conditions. Thus the theory remains
non-trivial, just its degrees of freedom are
actually those of a field on the boundary --
providing a non-trivial realization of holography
idea.

This example can be easily extended to more general
space-times with boundaries and boundaries can consist of
the stable points of other $Z_2$ transformations.\footnote{
In (\ref{bulk}) it is the symmetry $z_0 \rightarrow -z_0$,
which in the case of $d+1=2$ with complex $z = z_1 + iz_0$
becomes complex conjugation $z \rightarrow \bar z$.
Of considerable interest in the same dimension is the
case when the role of space-time is played by a
{\it hyperelliptic} Riemann surface $y^2(z) =
{\rm Polynomial\ of}\ z$, while the discrete
transformation is different: $y(z) \rightarrow -y(z)$.
In this example our "unphysical", Neumann minus Dirichlet,
propagator is exactly the one that appears in the role of the
two-point function $\rho^{(0|2)}(z,z')$ in matrix model theory,
with pole not at coincident, but at the $Z_2$-reflected points
$z'=z^*$, lying at two different sheets of the surface,
see \cite{amm,Eyn} for recent presentations.
Actually, one can use in the theory (\ref{bulk})  the
ordinary, {\it physical}, propagator, but than the
interaction term becomes explicitly non-local:
made from powers of
$\frac{1}{2}(\chi(z) + \bar\chi(z^*))$ instead of $\chi(z)$.}
For Euclidean space $R^{d+1}$ {\it per se} the index $s_n=1$,
while for $$ s_n = (n-2)\Delta_- - 1$$
(\ref{bulk}) and (\ref{bound}) describe a sub-sector in the theory of
scalar fields $\phi_\pm$ of particular dimensions
$\Delta_\pm = \frac{d\pm 1}{2}$ in $AdS_{d+1}$,
with $\phi_{\Delta_-} = z_0^{\Delta_-}{\rm Re}\ \chi(z)$
and $\phi_{\Delta_+} = z_0^{\Delta_-}{\rm Im}\ \chi(z)$.
Among other things, this fact manifests itself in
{\it conformal invariance} of effective theory (\ref{bound}).
Amusingly, even for the $AdS_{d+1}$ case, $s_n$ can sometime take value $1$:
this happens when interaction shape is adjusted to space-time
dimension so that $(n-2)((d+1)-2) = 4$, i.e. when
$(d+1,n) = (6,3),\ (4,4)$ or $(3,6)$.
To avoid possible confusion, we emphasize that
the action (\ref{bulk}) is essentially complex, still
most amplitudes are real, but the unitarity of the theory
(inessential for our considerations) can be under question.

In the remaining part of this paper we briefly
comment on straightforward derivations
of (\ref{bulk}) from the scalar theory in Euclidean $AdS_{d+1}$
and of (\ref{bound}) from (\ref{bulk}).

\section{Propagators in AdS}

\subsection{Action \label{acti}}

The free action of real scalar field $\phi_{\Delta}$ in $AdS_{d+1}$
with Euclidean metric $$ds^{2} =
\frac{dz_0^2 + d\vec z^2}{z_0^2}$$
is given by
$$
S = \frac{1}{2}\int_{AdS_{d+1}}
\sqrt{g}\left(g^{\mu\nu}\partial_\mu\phi_\Delta
\partial_\nu\phi_\Delta + m_\Delta^2\phi_\Delta^2\right)
 d^{d+1}z + S_{bound} = \phantom{5555}
$$\be
= \frac{1}{2}\int_{AdS_{d+1}} \frac{d^{d+1}z}{z_0^{d-1}}\left(
(\partial_0 \phi_\Delta)^2 + (\vec\partial \phi_\Delta)^2 +
\frac{m_\Delta^2\phi_\Delta^2}{z_0^2}\right) + S_{bound},
\label{AdSact}
\ee
where the AdS mass is related to AdS dimension $\Delta$ of
$\phi_\Delta$ through $$m_\Delta^2 = \Delta(\Delta-d).$$ Dimension $\Delta$
is restricted to $\Delta>\frac{d-2}{2}$ by the unitarity bound
\cite{Kl-Wit,Mez-Tow}.
This action can be transformed to the one in $R^{d+1}_+$ by
rescaling of field variables\footnote{
We thank V.Rubakov for useful discussions of this issue.}
$\phi_\Delta(z) = z_0^{\Delta_-}\chi_\Delta(z)$:
\be
S = \frac{1}{2}\int_{R^{d+1}_+} d^{d+1}z \left(
(\partial_0 \chi_\Delta)^2 + (\vec\partial \chi_\Delta)^2 +
\frac{m_\Delta^2+\Delta_+\Delta_-}{z_0^2}\chi_\Delta^2\right),
\label{Ract}
\ee
provided in (\ref{AdSact})
$$
S_{bound} = \frac{\Delta_-}{2} \oint_{\partial \Big(AdS_{d+1}\Big)}
\frac{d^d\vec z}{z_0^d}\ \phi_\Delta^2 =
\frac{\Delta_-}{2} \oint_{R^d} d^d\vec x
\lim_{z_0 \rightarrow +0}\frac{\chi_\Delta^2(z_0,\vec x)}{z_0}
$$
For special dimensions $\Delta = \Delta_\pm$ the AdS masses
$m_{\Delta_\pm}^2 = -\Delta_+\Delta_-$,
so that the last term at the r.h.s.
of (\ref{Ract}) vanishes,\footnote{
Potentially solvable (rational Calogero) situation arises for
entire integer-labeled tower of "dressed" masses:
$M^2_\Delta= \Delta(\Delta-d) + \Delta_+\Delta_- = N(N+1)$,
i.e. for $\Delta = \frac{1}{2}\Big(d \pm (2N+1)\Big)$. We do not discuss this
 -- as many other -- obvious
generalizations here.
}
and the action converts into a
free massless  action
for the scalar field in $R^{d+1}$:
\be
S = \frac{1}{2}\int_{R^{d+1}_+} d^{d+1}z \left(
(\partial_0 \chi_{\Delta_{\pm}})^2 + (\vec\partial \chi_{\Delta_{\pm}})^2\right)
\label{Rml}
\ee
  One more peculiar feature  of this particular choice of $\Delta$ is that
precisely at $\Delta = \Delta_\pm$ the scalar theory in the bulk
acquires extended symmetry: {\it local} conformal invariance.
Indeed, since the scalar curvature $R(z)$ of {\it conformal}
metric $g_{\mu\nu}(z) = \rho(z)\delta_{\mu\nu}$ in $d+1$
dimensions is equal to
$$
R = - \frac{d}{\rho} \left(\partial^2\log\rho\ +\ \frac{d-1}{4}\
\partial_\mu \log\rho\ \partial_\mu \log\rho\right),
$$
it follows that
$$
\frac{1}{2}\int \sqrt{g}\left( g^{\mu\nu} \partial_\mu\phi_{\Delta_{\pm}}
\partial_\nu\phi_{\Delta_{\pm}}
+ \xi R\phi_{\Delta_{\pm}}^{2}\right) d^{d+1}z
$$
is invariant under the simultaneous change
$$\left\{\begin{array}{c}
\rho(z) \longrightarrow \lambda^2(z)\rho(z), \\
\phi_{\Delta_{\pm}}(z) \longrightarrow
\lambda^{-\Delta_-}(z)\phi_{\Delta_{\pm}}(z)
\end{array}\right.
$$
with arbitrary $z$-dependent $\lambda(z)$,
satisfying $\partial_{0}\lambda(z_{0}=0)=0$  provided
$$\xi = \frac{d-1}{4d} = \frac{\Delta_-}{2d}$$
(for $d+1=4$ this $\xi = \frac{1}{6}$).
For $AdS_{d+1}$ our $\rho = \frac{1}{z_0^2}$ and $R = -d(d+1)$,
so that $\xi R(z) = - \frac{(d-1)(d+1)}{4} = - \Delta_-\Delta_+$,
i.e. exactly equals $m_{\Delta_{\pm}}^2$.
 Interactions preserve conformal symmetry, both local
and global, whenever $s_n = 1$. An interesting question is
how the {\it local} symmetry is realized in these cases
in the boundary theory (\ref{bound}).

\subsection{Bulk-to-boundary propagators}

Whenever the boundary is Euclidean space -- as is
the case with our parametrization of AdS -- it
is practical to use {\it mixed} representation
for Feynman diagrams: {\it momentum} along the boundary
and {\it coordinate} in the orthogonal direction (inside
the bulk).
Fourier transform of the boundary variable $\vec x$
converts the AdS bulk-to-boundary propagator \cite{Witt}
\be
\tilde K_\Delta(w,\vec x) = \frac{w_0^\Delta}
{\left[w_0^2 + (\vec w - \vec x)^2\right]^\Delta}
\label{butobo}
\ee
into\footnote{Schwinger parametrization is used to deal with the denominator:
$$\frac{1}{P^{\alpha}}=\frac{1}{\Gamma(\alpha)}\int\limits_{0}^{\infty}
d\lambda\ \lambda^{\alpha-1}e^{-\lambda P} $$
and the emerging integral is
$$ \int\limits_{0}^{\infty}\lambda^{\nu-1}
e^{-\alpha\lambda-\frac{\beta}{\lambda}}d\lambda =
2\left(\frac{\beta}{\alpha}\right)^{\frac{\nu}{2}}{\cal
K}_{\nu}(2\sqrt{\alpha\beta})$$
}
\be
K_\Delta(w|\vec p) = \int d^d\vec x e^{i\vec p \vec x}
\frac{w_0^\Delta}
{\left[w_0^2 + (\vec w - \vec x)^2\right]^\Delta}  =
w_0^\Delta e^{i\vec p \vec w}
\left(\frac{\vec p\ ^2}{w_0^2}\right)^{\frac{1}{2}
\left(\Delta - \frac{d}{2}\right)}
{\cal K}_{\Delta-\frac{d}{2}}\left(w_0\sqrt{\vec p\ ^2}\right)
\label{KthrcalK}
\ee
Here and below we systematically omit inessential factors to avoid
overloading the formulas. Equalities are defined modulo such factors.

For semi-integer index $\Delta-\frac{d}{2}$ the modified
Bessel function ${\cal K}$ turns into elementary function,
especially simple for particular values of
$\Delta = \Delta_\pm = \frac{1}{2}(d\pm 1)$.
This follows from the intermediate formulas:
$$
K_\Delta(w|\vec p)\ =\ w_0^\Delta \int d^d\vec x e^{i\vec p \vec x}
\int_0^\infty e^{-\alpha(w_0^2+(\vec w-\vec x)^2)} \alpha^{\Delta-1}
d\alpha \ =\ w_0^\Delta e^{i\vec p \vec w}\int_0^\infty e^{-\alpha
w_0^2 - \frac{|p\,|^2}{4\alpha}} \alpha^{\Delta-\frac{d}{2}-1}d\alpha\
= $$ \be =\ w_0^\Delta e^{i\vec p \vec w}\int_0^\infty e^{-\beta^2
w_0^2 - \frac{|p\,|^2}{4\beta^2}} \beta^{2\Delta-d-1}d\beta \ =\
w_0^\Delta e^{i\vec p \vec w}\int_0^\infty e^{
-\frac{1}{4}\lambda^2|p\,|^2 - \frac{w_0^2}{\lambda^2}}
\lambda^{d-2\Delta-1}d\lambda \ee Here $\alpha = \beta^2 =
\lambda^{-2}$ and $|p|=\sqrt{(\ \vec p\ )^2}$. Distinguished are values of
$\Delta$ when the powers of either $\beta$ or $\lambda$ disappear
from the pre-exponents, i.e. when $\Delta =  \Delta_\pm$. The
integral over $\beta$ (or $\lambda$) produces a factor of $w_0^{-1}$
(or $|p\,|^{-1}$) in the pre-exponent,\footnote{Here the celebrated
integral is used:
 $$\int_{0}^{\infty}e^{-a\lambda^2-\frac{b}{\lambda^{2}}}d\lambda
=\sqrt{\frac{\pi}{4a}}\ e^{-2\sqrt{ab}}$$}
and we get:
\be
K_{ \Delta_+}(w|\vec p) = w_0^{ \Delta_-}
e^{i\vec p\vec w} e^{-|p\,|w_0}
\label{K+}
\ee
(since $ \Delta_+ - 1 =  \Delta_-$), and
\be
K_{ \Delta_-}(w|\vec p) =
\frac{1}{|p\,|}w_0^{ \Delta_-} e^{i\vec p\vec w} e^{-|p\,|w_0} =
\frac{1}{|p\,|}K_{ \Delta_+}(w|\vec p).
\label{K-}
\ee
Eq.(\ref{K-}) implies that the amplitudes
with external fields of dimension $\Delta =  \Delta_-$ at the boundary
can be obtained from those of $\Delta =  \Delta_+$ by insertion of
appropriate $|p\,|^{-1}$ factors.

\subsection{Bulk-to-bulk propagators}
Propagators in the space-time with boundary depend on boundary conditions.
 The
natural physical requirement is that the flow
$\chi_{\Delta}\partial_{0}\chi_{\Delta}$ of field $\chi_{\Delta}$
 in (\ref{Ract})
through the boundary vanishes \cite{Rub}. This means that only Dirichlet or Neumann
boundary conditions can be imposed on the field $\chi_{\Delta}$ at $z_{0}=0$,
 and our theory can be quantized for each value of $\Delta$,
corresponding to the same value of mass. Thus, the propagator of the scalar
 field $\phi_\Delta$ in
$AdS_{d+1}$ satisfies
\be
\Big(-\Box_{AdS}(w) + m^{2}_\Delta \Big)G_\Delta(w,z)
= \frac{1}{\sqrt{g(z)}}\ \delta^{(d+1)}(w-z)
= (z_0)^{d+1}\delta^{(d+1)}(w-z)
\label{eqpro}
\ee
with additional requirement that
\be
G_\Delta(w,z) =
w_0^\Delta \ \ \ {\rm as} \ \ \ w_0 \rightarrow +0.
\label{asym}
\ee
Our consideration in subsection \ref{acti} implies that
for particular values of $\Delta = \Delta_\pm$
this propagator is -- modulo the factor of
$(w_0z_0)^{\Delta_-}$ -- the massless scalar propagator
in $R^{d+1}$, i.e. Fourier transform of
$\Big(p\,_0^2 + \vec p\ ^2\Big)^{-1}$. The difference between
$\Delta_+$ and $\Delta_-$ is in the boundary condition (\ref{asym}):
$$
G_{\Delta_+}(w,z) = (w_0z_0)^{\Delta_-}\!\!\int \frac{d \vec p\ d p_{0}}{p_0^2+\vec p\,^2}\
e^{i\vec p\,(\vec w-\vec z)} \sin (p_0z_0) \sin (p_0w_0) \ =
\vspace{-0.5cm}
$$\be
=\ (w_0z_0)^{\Delta_-}\left\{\frac{1}{\Big((w_0-z_0)^2 +
(\vec w - \vec z)^2\Big)^{\Delta_-}}\ - \ \frac{1}{\Big((w_0+z_0)^2 +
(\vec w - \vec z)^2\Big)^{\Delta_-}}\right\} \ =\
\frac{1}{u^{\Delta_-}}\ - \ \frac{1}{(u+2)^{\Delta_-}}
\label{pro+}
\ee
satisfies Dirichlet, while
$$
G_{\Delta_-}(w,z) = (w_0z_0)^{\Delta_-}\!\!\int
\frac{d \vec p\ d p_{0}}{p_0^2+\vec p\,^2}\
e^{i\vec p\,(\vec w-\vec z)} \cos (p_0z_0) \cos (p_0w_0) \ =
\vspace{-0.5cm}$$\be
=\ (w_0z_0)^{\Delta_-}\left\{\frac{1}{\Big((w_0-z_0)^2 +
(\vec w - \vec z)^2\Big)^{\Delta_-}}\ + \ \frac{1}{\Big((w_0+z_0)^2 +
(\vec w - \vec z)^2\Big)^{\Delta_-}}\right\} \ =\
\frac{1}{u^{\Delta_-}}\ + \ \frac{1}{(u+2)^{\Delta_-}}
\label{pro-}
\ee
-- Neumann boundary conditions at the boundary of $AdS$ (at $w_0 =0$
or $z_0=0$).
Integrals  in (\ref{pro+}) and (\ref{pro-})
are evaluated by introduction of auxiliary integration,
$$\frac{1}{p\,_0^2+\vec p\ ^2} = \int_0^\infty e^{-\alpha(p\,_0^2+\vec p\,^2)}
d\alpha,$$ followed by Gaussian integration over $p\,_0$ and $\vec p$.
Switching to $\lambda = (4\alpha)^{-1}$ we obtain:
$$
G_{\Delta_+}(w,z)\ =\ (w_0z_0)^{\Delta_-}\int_0^\infty
\lambda^{\Delta_--1}d\lambda \left(
\exp\left[-\lambda\Big((w_0-z_0)^2 + (\vec w-\vec z)^2\Big)\right] -
\exp\left[-\lambda\Big((w_0+z_0)^2 + (\vec w-\vec z)^2\Big)\right]
\right)
$$
and integral over $\lambda$ provides (\ref{pro+}).
The last representations in formulas (\ref{pro+}) and (\ref{pro-})
are in terms of the usual AdS variables
$$u = \frac{(w_0-z_0)^2 + (\vec w-\vec z)^2}{2w_0z_0}\ \ \ \
{\rm and} \ \ \ \
u+2 = \frac{(w_0+z_0)^2+(\vec w-\vec z)^2}{2w_0z_0}$$

For the sake of completeness,
in Appendix A  at the end of the paper, we present alternative
derivation of these propagators: as solutions of the hypergeometric
equation.

\section{The difference of AdS propagators and the theory (\ref{bulk})}

Now we are ready to introduce our simplified AdS theory.
Suggestion is to substitute the propagators (\ref{pro-}) and
(\ref{pro+}) by their peculiar linear combination: subtract
one from another and define
$$
G_0(w,z) \equiv \frac{1}{2}\Big(G_{\Delta_-}(w,z) -
G_{\Delta_+}(w,z)\Big) =
(w_0z_0)^{\Delta_-}\!\!\int \frac{d \vec p\ d p_{0}}{p\,_0^2+\vec p\,^2}
e^{i\vec p\,(\vec w-\vec z)} \cos \Big(p\,_0(w_0+z_0)\Big) \ =
$$ $$ =\ \frac{1}{(u+2)^{\Delta_-}} =
\left(\frac{w_0z_0}{(w_0+z_0)^2+(\vec w-\vec z)^2}\right)^{\Delta_-}
$$
as the propagator of a new conformally invariant scalar field theory.
The spatial Fourier transform of such bulk-to-bulk propagator is especially
simple:
\be
G_0(w,z) = (w_0z_0)^{\Delta_-}\int  \frac{d^d\!\vec p}{|p\,|}\
e^{i\vec p\,(\vec w-\vec z)} e^{-|p\,|(w_0+z_0)}
\label{prod}
\ee
The fact that $z_0$ and $w_0$ appear in the exponent as a simple sum\footnote{
Unlike they show up in Fourier transform of the {\it physical}
propagator $\ {\cal G} = \frac{1}{2}(G_{\Delta_-}+G_{\Delta_+})\
\sim\ 1/u^{\Delta_-}$, $\ $
which contains a far more
sophisticated factor of $\ e^{i|p\,||w_0-z_0|}$, see appendix C below.},
makes convolutions of propagators in expressions for
Feynman diagrams very simple and leads to especial simplicity of
effective theory (\ref{bound}).

\begin{figure}[h]
\centerline{\includegraphics[width=100mm]{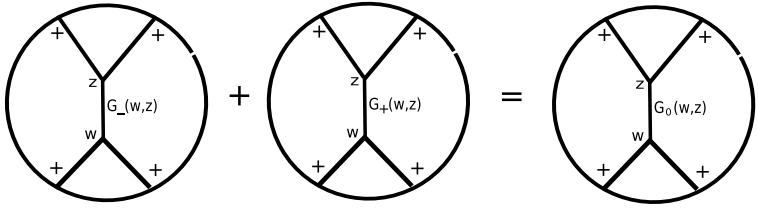}}
\caption{{\footnotesize  Four-point function. The contribution of $G_{\Delta_-}(w,z)$
and $G_{\Delta_+}(w,z)$ in the bulk gives propagator $G_0(w,z)$}}
\label{fig8}
\end{figure}

What is the way to realize such projection -- from two different
propagators to their difference? A possible answer is provided by the
theory of two real scalar fields $\phi_{\Delta_-}$ and $\phi_{\Delta_+}$,
subjected to Neumann and Dirichlet boundary conditions respectively with the
action
$$
S = \int d^{d+1}z
\sqrt{g}\left[\frac{1}{2}g^{\mu\nu}\partial_\mu\phi_{\Delta_-}
\partial_\nu\phi_{\Delta_-} + \frac{1}{2}m_{\Delta_{\pm}}^2\phi_{\Delta_-}^2+
\frac{1}{2}g^{\mu\nu}\partial_\mu\phi_{\Delta_+}
\partial_\nu\phi_{\Delta_+} +
\frac{1}{2}m_{\Delta_\pm}^2\phi_{\Delta_+}^2 +
\sum\limits_{n}\frac{t_n}{n!}(\phi_{\Delta_-}\!\!+i\phi_{\Delta_+})^{n}
\right]
$$
To prove that this theory reproduces the bulk-to-bulk propagator
 $G_{0}(w,z)$, consider the simplest case, when only $t_{3}\neq 0$ in
the interaction terms. In the figure \ref{fig8}, the tree contribution
to the four-point function of $\phi_{\Delta_+  }$ is shown.
\bigskip
\bigskip
There are two diagrams: one with the bulk-to-bulk propagator $G_{\Delta_-}(w,z)$,
another one with $G_{\Delta_+}(w,z)$. The vertex contribution for
$\phi_{\Delta_+}^{3}$ interaction term is $V_{+++}=i^{3}t_{3}\sqrt g$, while
for $\phi_{\Delta_+}^{2}\phi_{\Delta_-}$ is $V_{++-}=i^{2}t_{3}\sqrt g$. Thus,
$V_{+++}=iV_{++-}$. This extra $i$ factor, being squared (since there are two vertices
in each diagram), gives relative {\it minus} sign between these two diagrams.
Hence, effectively, the propagator in this theory is
$G_{\Delta_-}(w,z)-G_{\Delta_+}(w,z)=G_{0}(w,z)$. The generalization to
other tree and loop  diagrams is straighforward.

In terms of fields
\begin{eqnarray}
\nonumber
\phi=\phi_{\Delta_-}\!+i\phi_{\Delta_+}\\
\nonumber
\overline{\phi}=\phi_{\Delta_-}\!-i\phi_{\Delta_+}
\end{eqnarray}
this action can be rewritten as
\begin{equation}
\label{SAdS}
S=\int d^{d+1}z\sqrt{g}\left
(\frac{1}{2}g^{\mu\nu}\, \partial_{\mu}\phi \partial_{\nu}\overline{\phi}\
+\ \frac{m_{\Delta_{\pm}}^{2}}{2}\phi\overline{\phi}\  +\
 \sum\limits_{n}\frac{t_{n}}{n!}\phi^{n}\right).
\end{equation}
Making rescaling of variables
\be
\nonumber
\phi_{\Delta_-}=z_{0}^{\Delta_-}\chi_{\Delta_{-}}\\
\nonumber \phi_{\Delta_+}=z_{0}^{\Delta_-}\chi_{\Delta_{+}} \ee
from the section 2.1, we obtain, up to boundary terms, the action
\be S=\int d^{d+1}z
\left[\frac{1}{2}\partial_{\mu}\chi\partial_{\mu}\overline{\chi}\
+\ \sum\limits_{n}\frac{t_n}{n!}z_{0}^{s_n-1}\chi^{n}\right] \ee
for the fields $\chi=\chi_{\Delta_-}\!+i\chi_{\Delta_+}$ and
$\overline{\chi}=\chi_{\Delta_-}\!-i\chi_{\Delta_+}$. Index
$s_n=(n-2)\Delta_{-}-1$.

\bigskip
\bigskip
\bigskip

\section{Tree diagrams}
In this section we evaluate arbitrary tree correlators
of the fields of dimension $\Delta_+$ on the boundary of
$AdS_{d+1}$ in the complex-scalar field theory (\ref{bulk}),
or equivalently (\ref{SAdS})
with {\it holomorphic} interaction.  
It turns out that all the tree diagrams are expressed
as simple rational functions of $d$-dimensional (rather than
$d+1$-dimensional) momenta: one extra dimension can be
explicitly integrated out and provide a non-local, but
rather simple effective theory (\ref{bound}).
Moreover, the same effective theory appears to describe all
loop diagrams as well.

\subsection{Single-vertex diagram
\label{svd}}

\begin{figure}[h!]
\centerline{\includegraphics[width=42mm]{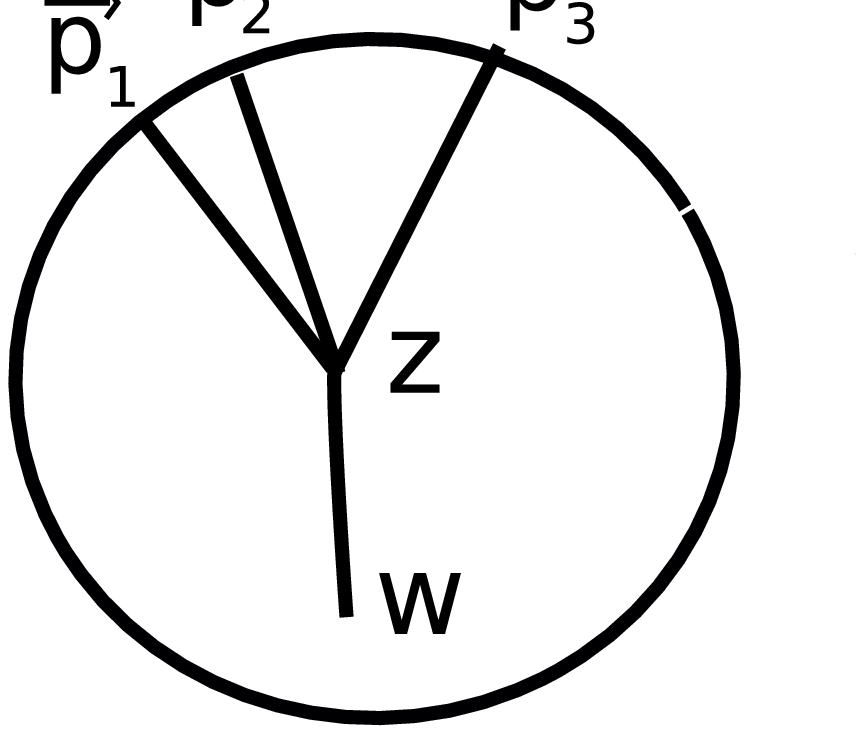}}
\caption{{\footnotesize One-vertex diagram, $w$ is arbitrary
point in the bulk. Four-valent vertex is chosen for illustrative
purposes.}}
\label{fig1}
\end{figure}

We begin from the simplest diagram with a single vertex of
valence $n+1$, see Fig.\ref{fig1}.
\noindent
The corresponding amplitude is given by
$$
A(w|\vec p_1,\ldots,\vec p_n) = t_{n+1}\int\frac{dz_0d^d\vec z}{z_0^{d+1}}
\ G_0(w,z) K_{\Delta_+}(z|\vec p_1)\ldots K_{\Delta_+}(z|\vec p_n) \
\stackrel{(\ref{prod}) \& (\ref{K+})}{=}\
$$ $$=\
t_{n+1}\int\frac{dz_0d^d\vec z}{z_0^{d+1}}
\left((w_0z_0)^{\Delta_-}\int  \frac{d\vec p}{|p\,|}
\ e^{i\vec p\,(\vec w-\vec z)}
e^{-|p\,|(w_0+z_0)}\right)
\left(z_0^{ \Delta_-} e^{i\vec p_1\vec z} e^{-|p_1|z_0}\right)\ldots
\left(z_0^{ \Delta_-} e^{i\vec p_n\vec z} e^{-|p_n|z_0}\right)
\ = \vspace{-0.3cm}$$\be =\
t_{n+1}\frac{w_0^{\Delta_-}}
{|p_{1n}|\Big(|p_1|+\ldots + |p_n|+|p_{1n}|\Big)^{s_{n+1}}}\
e^{i\vec w (\vec p_1 + \ldots + \vec p_n)}e^{-|p_{1n}|w_0}
\label{A2++}
\ee
where $|p_{1n}| \equiv \sqrt{(\vec p_1+ \ldots + \vec p_n)^2}$.
Note, that the use of projection (\ref{prod}) instead of (\ref{pro+})
or (\ref{pro-}) eliminates the poles at collinear momenta, when
at $|p_{1n}| = |p_1| + \ldots + |p_n|$, which normally occur
in AdS Feynman diagrams, see Appendix $C$ below.

\subsection{Multi-vertex diagrams \label{mvd}}

Comparison of (\ref{A2++}) and (\ref{K+}) shows that their
$w$-dependencies are exactly the same. This implies {\it universality}
of the vertex insertion and the possibility of recursive evaluation
of tree diagrams. Namely we can factor out all the $w$ dependence in
a simple and universal manner:
\be
A^\Gamma(w|\vec p_1,\ldots,\vec p_{n}) = w_0^{\Delta_-}
e^{i\vec w(\vec p_1 + \ldots + \vec p_n)}
e^{-w_0|\vec p_1 + \ldots + \vec p_n|}
B^\Gamma(\vec p_1,\ldots,\vec p_{n}) =
K_{\Delta_+}(w|\vec p_1 + \ldots + \vec p_n)
B^\Gamma(\vec p_1,\ldots,\vec p_{n}),
\label{factoriz}
\ee
where $\Gamma$ labels the graph = diagram,
which in our case is the rooted tree
with the total momentum
$\vec p\,_\Gamma \equiv \vec p_1 + \ldots  + \vec p_n$.

\begin{figure}[h]
\centerline{\includegraphics[width=82mm]{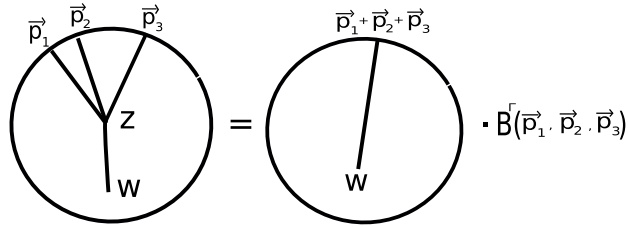}}
\caption{{\footnotesize First step. Integration over $d^{d+1}z$ with
 the weight $G_{0}(w,z)$ converts the bunch of bulk-to-boundary propagators
into a single bulk-to-boundary propagator.}}
\label{fig2}
\end{figure}

Eq.(\ref{factoriz}) is pictorially represented in Fig.\ref{fig2}.
It shows that in the theory (\ref{bulk}) or equivalently (\ref{SAdS})
any number of bulk-to-boundary propagators,
attached to an intermediate vertex at $z$ in the bulk,
can after integration over $z$
be substituted by a single bulk-to-boundary propagator,
leading to the point $w$ in the bulk and carrying the total
along-the-boundary momentum, multiplied by the factor
$B^\Gamma(\vec p_1,\ldots,\vec p_{n})$.

\bigskip

\begin{figure}[h!!!]
\centerline{\includegraphics[width=30mm,angle=-90]{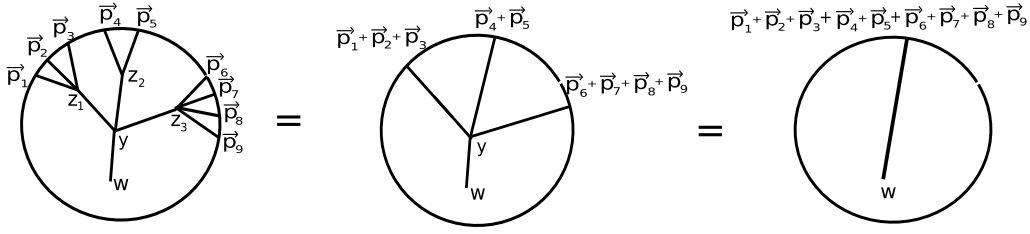}}
\caption{{\footnotesize Second step. Factors $B^{ƒ}(\vec p_{i})$ omitted.}}
\label{fig3}
\end{figure}

At the next step we evaluate the second-level diagram like
the left one in Fig.\ref{fig3} with several single-vertex
sub-diagrams meeting at the single vertex at point $y$ in the
bulk. As shown in the picture, repeated application of
above-described contraction provides an amplitude, again
proportional to a single bulk-to-boundary propagator.
\bigskip
\bigskip
We can apply this algorithm as many times as necessary,
converting arbitrary tree diagram with the root at $w$ inside
the bulk into a single bulk-to-boundary propagator
$K_{\Delta_+}(w|{\rm total\ momentum})$. The coefficient function
arises from recurrent relation
\be
B^\Gamma = t_{n+1}
\frac{B^{\Gamma_1}\ldots B^{\Gamma_n}}
{|p\, _\Gamma|\Big(|p\, _{\Gamma_1}|
+ \ldots + |p\, _{\Gamma_n}| + |p\, _{\Gamma}|\Big)^{s_{n+1}}}
\ee
From this relation it is clear, that induced diagrammatic technique, is {\it
local}: each vertex contribution depends on momenta,
incoming into this particular vertex only. This {\it locality} is due to
specific choice of propagator $G_{0}(w,z)$.

\begin{figure}[h]
\centerline{\includegraphics[width=32mm]{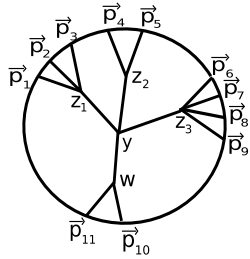}}
\caption{{\footnotesize The last step. Emission of the bunch
of bulk-to-boundary
propagators from the final vertex at the point $w$ in the bulk.}}
\label{fig4}
\end{figure}

To finish evaluation of the diagram in Fig.\ref{fig4}
it remains to attach the last bunch of bulk-to-boundary propagators
to the root $w$ of the tree (Fig.\ref{fig3} left) and integrate over $w$.
Such convolution of $m$ propagators is equal to:
\be
\int K_{\Delta_+}(w|\vec p_1)\ldots K_{\Delta_+}(w|\vec p_m)
\frac{dw_0 d\vec w}{w_0^{d+1}} =
\frac{\delta(\vec p_1 + \ldots + \vec p_m)}
{\Big(|p_1| + \ldots + |p_m|\Big)^{s_m}}
\ee
Following this procedure one can explicitly evaluate any particular
tree diagram.

\bigskip

\subsection{Feynman rules and effective action}

 The  result of our consideration can be formulated as simple
Feynman rules for the $B^{\Gamma}$ functions:
$$
\begin{tabular}{|c|c|}
\hline &\\
\ propagator (line)\ &  \ \
$\frac{1}{|p\,|}$ \\
&\\ \hline &\\
$n$-vertex &
$t_n\frac{\delta^{(d)}(\vec p_1 + \vec p_2 + \ldots + \vec p_n)\phantom{5_5}}
{\Big(|p_1|+|p_2|+\ldots +|p_n|\Big)^{s_n\phantom{5^5}\hspace{-0.4cm}}}$\\
&\\ \hline
\end{tabular}
$$
External lines carry the same factors $|p\,|^{-1}$ provided external
fields have dimension $\Delta_-$, while for dimension $\Delta_+$
external lines carry no factors of momentum. Moreover,  as explained in
section 5 below, the same rules work for loop calculations.

This diagram technique can be summarized in the form of a simple effective
action: tree and loop diagrams for scalars of dimension $\Delta_{\pm}$
at the boundary and peculiar propagator $G_{0}(w,z)$
in the bulk coincide with the
tree and loop diagrams in the {\it boundary} theory (\ref{bound}):
\be
{\cal Z}\{J\} =
\int D\varphi(\vec x) \exp \left\{
-\frac{1}{g^2}\oint d\vec x \left(\frac{1}{2}\varphi \sqrt{\Box} \varphi +
\Big(J_N+J_D\sqrt{\Box}\Big)\varphi +
\sum_n \frac{t_n}{n!} \int_0^\infty
\left(e^{-\alpha\sqrt{\Box}}\varphi\right)^n
\alpha^{s_n-1} d\alpha\right)\right\}
\label{ZADS}
\ee

This theory is conformal invariant: rescalings of field $\varphi$,
which has dimension $\Delta_-$ are accompanied by the transformation of
auxiliary integration variable $\alpha$, which has dimension $-1$.
In variance with the boundary models, usually considered in the context
of AdS/CFT correspondence
\cite{AdS/CFT}-\cite{Oz}, \cite{Gopak}, \cite{Rast}-\cite{Akh},
the theory (\ref{ZADS}) is
non-local. Also trees (loops) in the bulk are the same trees (loops) in the
theory (\ref{ZADS}).

We emphasize that matching between theories (\ref{bulk}) and (\ref{ZADS}) is
valid for arbitrary value of index $s_n$
(not restricted to $s_n=(n-2)\Delta_--1$).
Only for particular choice of $s_n=(n-2)\Delta_--1$, the bulk theory (\ref{bulk})
describes the AdS theory (\ref{SAdS}).

\bigskip

\section{Loops \label{loops}}

It is straightforward to see that our Feynman rules -- and thus the
effective theory (\ref{bound}) at the boundary  -- reproduce expressions
for loop diagrams in AdS theory. Again, in variance with the usual AdS/CFT
correspondence, loops in (\ref{SAdS}) are loops in (\ref{ZADS}).

\subsection{Sample 1-loop diagram.}

\begin{figure}[h]
\centerline{\includegraphics[width=30mm,angle=0]{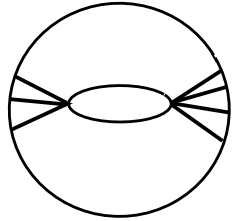}}
\caption{{\footnotesize One loop diagram.}}
\label{fig5}
\end{figure}

We provide just an illustration. Consider
the diagram in Fig.\ref{fig5}.
\noindent
Original expression is
\be
t_nt_m
\int\int \frac{dz_0d^d\vec z}{z_0^{d+1}}\frac{dw_0d^d\vec w}{w_0^{d+1}}
K_{\Delta_{+}}(z|\vec p_1)\ldots K_{\Delta_{+}}(z|\vec p_{n-2}) G_0^2(z,w)
K_{\Delta_{+}}(w|\vec q_1)\ldots K_{\Delta_{+}}(w|\vec q_{m-2})
\label{orex}
\ee
According to (\ref{K+}) and (\ref{prod}) the product of propagators
is equal to
$$
z_0^{\Delta_{-}(n-2)}e^{i\vec z\sum_{i} \! \vec p_i}e^{-z_0\!\sum_{i}\! |p_i|}
\left(
(z_0w_0)^{\Delta_-}\int \frac{d^d\vec r}{|r|}\ e^{i\vec r(\vec z-\vec w)}
e^{-|r|(w_0+z_0)}\right)^2
w_0^{\Delta_{-}(m-2)}e^{i\vec w\sum_j\! \vec q_j}e^{-w_0\!\sum_j\! |q_j|}
$$
and after integration over $\vec z$ and $\vec w$ (\ref{orex}) turns into
$$ t_nt_m
\int\int \frac{d^d\vec r}{|r|}\frac{ d^d\vec r\,^\prime}{|r^\prime|}
\delta^{(d)}\Big(\sum_i \vec p_i +\vec r +\vec r\,^\prime\Big)
\delta^{(d)}\Big(\sum_j \vec q_j -\vec r -\vec r\,^\prime\Big)\times$$ $$\times
 \int_0^\infty z_0^{(n-2)\Delta_{-}-2}dz_0
\int_0^\infty w_0^{(m-2)\Delta_--2}dw_0
e^{-(\sum_i |p_i| + |r| + |r^\prime|)z_0}
e^{-(\sum_j |q_j| + |r| + |r^\prime|)w_0}
=$$ $$= t_nt_m\delta^{(d)}\left(\sum_i \vec p_i + \sum_j \vec q_j\right)
\int \frac{d^d \vec r}{|r||r^\prime|
\Big(\sum_i |p_i| + |r| + |r^\prime|\Big)^{s_n}
\Big(\sum_j |q_j| + |r| + |r^\prime|\Big)^{s_m}}
$$
where in the last formula
$|r^\prime| = \sqrt{\Big(\sum_i \vec p_i + \vec r\,\Big)^2} =
\sqrt{\Big(\sum_j \vec q_j - \vec r\,\Big)^2}$.
This expression is exactly the one which arises in effective theory (\ref{bound})
for the loop diagram of the same topology, Fig.\ref{fig5}, .

Generalizations to all other loop diagrams is straightforward.

\bigskip

\subsection{Tadpoles and divergencies}

\begin{figure}[h]
\centerline{\includegraphics[width=30mm,angle=0]{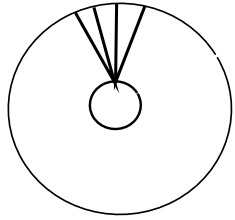}}
\caption{{\footnotesize Tadpole diagram.}}
\label{fig6}
\end{figure}

\begin{figure}[h]
\centerline{\includegraphics[width=30mm,angle=0]{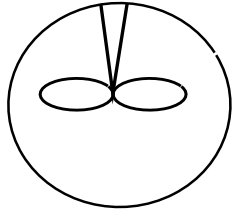}}
\caption{{\footnotesize Two loop diagram,
associated with the coupling $t_{6}$.}}
\label{fig7}
\end{figure}

A few comments are deserved by tadpole diagrams, like those shown in
Figs.\ref{fig6},\ref{fig7}. First of all, the projected propagator
$G_0(z,w)$ -- in variance from the usual ones, like $G_{\Delta_\pm}(z,w)$ --
is not singular at coincident points:
$$ G_0(z,z) = z_0^{2\Delta_-}\int \frac{d^d\vec p}{|p\,|}\ e^{-2|p\,|z_0} =
const $$ (this is also clear from its expression through $u$-variable,
since at $w=z$ this $u=0$, but $u+2 = 2\neq 0$). Thus the UV divergences
are absent, as one expects in the non-local theory (\ref{bound}) with
exponential damping of interactions at large momenta.

However, some peculiar divergences still survive.
For example, for the diagram in Fig.\ref{fig6} we have, according to
our Feynman rules:
\be
t_n\delta^{(d)}\left(\sum_{i=1}^{n-2} \vec p_i\right)
\int \frac{d^d \vec q}{|q|\Big(2|q| + \sum_{i=1}^{n-2}|p_i|\Big)^{s_n}}
\ =\ t_n\frac{\delta^{(d)}\left(\sum_{i=1}^{n-2} \vec p_i\right)}
{\left(\sum_{i=1}^{n-2}|p_i|\right)^{(n-4)\Delta_--1}}
\label{tadint}
\ee
More accurately, expression in the r.h.s. is valid only when
$(n-4)\Delta_- > 1$, otherwise the integral in the l.h.s.
diverges at large $|q|$. (Note, that vanishing sum of the space momenta
$\sum \vec p$ does not imply that the sum of their moduli, $\sum |p\,|$,
 is zero. This quantity is
always positive.) If we rewrite (\ref{tadint}) in coordinate
representation, making use of $G_0(z,z) = const$, we get:
$$
t_n\int \frac{dz_0d^d\vec z}{z_0^{d+1}}
K_{\Delta_{+}}(z|\vec p_1)\ldots K_{\Delta_{+}}(z|\vec p_{n-2})G_0(z,z) =
t_n\delta^{(d)}\left(\sum_{i=1}^{n-2} \vec p_i\right)
\int_{0}^{\infty}  e^{-z_0\sum_i |p_i|}z_0^{(n-4)\Delta_--2}dz_0
$$
In this representation it is clear that divergence, though looking like UV
from the point of theory (\ref{bound}), comes from the region
of small $z_0$, from the vicinity of the boundary of AdS and thus actually
an IR divergence.

Particular diagram in Fig.\ref{fig6} converges if $n\geq 6$, we consider
$d\geq 2$, and
divergence can be eliminated, say, by putting $t_3=t_4=t_5=0$. However,
the two-loop diagram in Fig.\ref{fig7}, associated with the coupling
$t_6$, diverges for exactly the same reason as the one loop in
Fig.\ref{fig6}.

\subsection{Coleman-Weinberg potential}
The boundary theory (\ref{bound}) can be used to sum up the one-loop
diagrams in the background of {\it constant} field $\Phi$.
The result of this summation -- the Coleman-Weinberg action for
our theory -- is\footnote{We use the standard background field technique.
The field $\varphi$ is decomposed into the sum of background and quantum
field: $\varphi=\Phi+\varphi_q$. We find the
part of the action proportional to $\varphi_{q}^{2}$, then diagonalize it in
momentum representation.
Arrow implies the use of relation $\log {\rm Det} = {\rm Tr} \log$
and subtraction of the $\Phi$-independent part from $S_{CW}(\Phi)$.
The one-loop contribution to $\log{\cal Z}\{J\}$
is related to $S_{CW}(\Phi)$ by Legendre transform.
}
\be
S_{CW}\{\Phi\} \sim\ \log {\rm Det}\left(\sqrt{\Box} +
\sum_n \frac{t_n \Phi^{n-2}}{(n-2)!}\frac{1}{(2\sqrt{\Box})^{s_n}}\right)\
\longrightarrow \
\int \log \left(1 + \sum_n \frac{t_n}{(n-2)!}
\frac{\Phi^{n-2}}{|q|(2|q|)^{s_n}}\right) d^d\vec q
\label{CWA}
\ee
For $s_n + 1 > d$, i.e. for $(n-4)\Delta_- > 1$ the integral converges
at large $|q|$ (otherwise there is our familiar IR divergence at small
$z_0$), and it converges for small $|q|$, though individual terms of
power expansion in $\Phi$ are divergent.

It is instructive to compare this Coleman-Weinberg action with the
naive open-sector boundary effective action, associated with
the free-field theory with quadratic vertex operators:
\be
{Z}_{op}\{I\} =
\int D\tilde\varphi(\vec x) \exp \left\{
-\frac{1}{g^2}\oint d^d\vec x
\left(\tilde\varphi {\Box} \tilde\varphi +
I\tilde\varphi^2\right)\right\} \ \sim\
{\rm Det}^{-1/2} \Big(\Box + I\Big)
\label{Zop}
\ee
For constant source $I(\vec x) = const$
\be
S_{op}(I) = \log {Z}_{op}\{I\} \ \sim\
\int  \log \left(1 + \frac{I}{q^2}\right)\ d^d\vec q
\ee
and does not have anything to do with $\log{\cal Z}\{J\}$. Still,
for distinguished value of index $s_n=1$ and for $I=\Phi^{n-2}$, it coincides
with $S_{CW}(\Phi)$ -- the one-loop contribution to Legendre transform of
$\log{\cal Z}\{J\}$ at constant $\Phi$.
Note also that the dimensions $\Delta = d-2$, considered in \cite{Gopak}
coincide with our $\Delta_\pm = \frac{d\pm 1}{2}$ for $d=3$ and $d=5$.

\section{Appendix A. Propagators as hypergeometric functions}

Usually in the literature eq.(\ref{eqpro}) is not solved explicitly
for particular dimensions $\Delta = \Delta_\pm$, as we did in this
paper. Instead one  uses the transcendental expression
for the bulk-to-bulk propagator through hypergeometric series
\cite{Rast,Kl-Wit,Gopak},
\be
G_\Delta(w,z) =
\frac{1}{(4\pi)^{\frac{d+1}{2}}}\frac{\Gamma(\Delta)\Gamma(\Delta-\Delta_-)}
{\Gamma(2\Delta-2\Delta_-)}
\left(\frac{2}{u}\right)^\Delta
F\left(\Delta,\ \Delta-\Delta_-,\ 2\Delta-2\Delta_-;\
-\frac{2}{u}\right)
\label{hyperpro}
\ee
In this section we briefly explain, how our
simple calculations are related to this standard approach.

For particular dimensions $\Delta = \Delta_\pm$,
 up to an overall normalization factor $
1/{(4\pi)^{\frac{d+1}{2}}}$,
we get\footnote{
Here we use the definition of Gauss hypergeometric series  $\phantom._2F_1$,
$$
F(a,b,c;z) = \frac{\Gamma(c)}{\Gamma(a)\Gamma(b)}\sum_{n=0}^\infty
\frac{\Gamma(n+a)\Gamma(n+b)}{n!\Gamma(n+c)}\ z^n, $$
which solves the hypergeometric equation,
$$
z(1-z)F^{\prime\prime} + \Big(c - (a+b+1)z\Big)F^\prime - abF = 0,
$$
and Newton's binomial expansion
$$
(1-z)^{-s} = \sum_{n=0}^\infty \frac{\Gamma(n+s)}{n!\Gamma(s)}\ z^n =
F(s,c,c;z)
$$}
from
(\ref{hyperpro}) :
$$
G_{\Delta_-} = \frac{2^{\Delta_-+1}\Gamma(\Delta_-)}{u^{\Delta_-}}\
F\left(\Delta_-,\ 0,\ 0;\ -\frac{2}{u}\right)
= \vspace{-0.3cm}$$ \be =
\frac{2^{\Delta_-}\Gamma(\Delta_{-})}{u^{\Delta_-}}\left( 1 +
\sum_{n=0}^\infty \frac{\Gamma(n+\Delta_-)}{\Gamma(\Delta_{-})n!}\left(-\frac{2}{u}\right)^n
\right) = 2^{\Delta_-}\Gamma(\Delta_-)
\left(\frac{1}{u^{\Delta_-}}+\frac{1}{(u+2)^{\Delta_-}}\right)
\label{prot-}
\ee
and
$$
G_{\Delta_+} = \frac{2^{\Delta_+}\Gamma(\Delta_+)}{u^{\Delta_+}}
F\left(\Delta_+,\ 1,\ 2;\ -\frac{2}{u}\right) =
\frac{2^{\Delta_+}}{u^{\Delta_+}}
\sum_{n=0}^\infty \frac{\Gamma(n+\Delta_+)\Gamma(n+1)}{n!\Gamma(n+2)}
\left(-\frac{2}{u}\right)^n =
$$
\be = -\frac{2^{\Delta_+}}{2u^{\Delta_-}}
\sum_{n=1}^\infty \frac{\Gamma(n+\Delta_-)}{n!}
\left(-\frac{2}{u}\right)^n
= 2^{\Delta_-}\Gamma(\Delta_-)
\left(\frac{1}{u^{\Delta_-}}-\frac{1}{(u+2)^{\Delta_-}}\right)
\label{prot+}
\ee

 An extra coefficient $2$ in the first line of (\ref{prot-}) comes from the
fact that the ratio $\frac{\Gamma(\Delta- \Delta_-)}
{\Gamma(2\Delta-2\Delta_-)}
\rightarrow 2$ as $\Delta \rightarrow \Delta_-$. For the same reason the
$n=0$ term in the sum enters with an extra coefficient $2$, this is taken
into account by an extra item $1$ in the first formula in the second line
of (\ref{prot-}).

Up to a common normalization factor 
the formulas (\ref{prot-}) and (\ref{prot+}) reproduce (\ref{pro-})
 and (\ref{pro+})
respectively.

\section{Appendix B. Derivations {\it a la} \cite{Rast}}

We illustrate this standard method by re-examining the example of
(\ref{A2++}), with $G_{0}(w,z)$ replaced by $G_{\Delta}(w,z)$.
We put $n=3$ (triple vertex) and $d=5$ to simplify the
formulas. The results of this appendix can be generalized to arbitrary
values of $n$ and $d$ in an obvious way.
Following \cite{Rast} we extract this quantity from solution of the
equation (\ref{eqpro}):
\be
(-\Box_{AdS} + m_{\Delta}^2) A_{\Delta}(w|\vec p_1,\vec p_2) =
K_{\Delta_\pm}(w|\vec p_1)K_{\Delta_\pm}(w|\vec p_2)\ \sim\
e^{-(|p_1|+|p_2|)w_0}e^{i(\vec p_1+\vec p_2)\vec w}w_0^{2 \Delta_-}
\label{eqn}
\ee
the subscript $\Delta$ for $A_{\Delta}(w|\vec p_1, \vec p_2)$ denotes
 dimension of the bulk-to-bulk propagator, eq.(\ref{ff})
below  restricts it to be $\Delta = \Delta_\pm$.
Proportionality sign in (\ref{eqn}) shows that the possible
$|p_1|^{-1}$ and $|p_2|^{-1}$ factors from $K_{\Delta_-}$ are ignored.
Substituting
$A_\Delta(w|\vec p_1,\vec p_2) = e^{i\vec w (\vec p_1 + \vec p_2)}
e^{-t\mu} F(t)$
with $t = w_0\sqrt{(\vec p_1 + \vec p_2)^2} = w_0|p_{12}|$ and $\mu =
\frac{|p_1|+|p_2|}{|p_{12}|}$, we obtain for $F(t)$:
$$
-t^2 F^{\prime\prime} + (2\mu t^2 + (d-1)t)F^\prime +
\left((1-\mu^2)t^2 - \mu(d-1)t + \Delta(\Delta - d)\right)F
= t^{2 \Delta_-}|p_{12}|^{-2\Delta_{-}}
$$
 Substituting further $F(t) = t^{\Delta_{-}} f(t)$, we get:
\be
- f^{\prime\prime} + 2\mu f^\prime + (1-\mu^2)f =
t^{-2+ \Delta_{-}}|p_{12}|^{-2\Delta_{-}}
\label{ff}
\ee
Generic solution of this equation is:
\be
A_{\Delta}(w|\vec p_1,\vec p_2) =
\frac{w_0^{\Delta_{-}}}{|p_{12}|^2 - (|p_1|+|p_2|)^2}\
e^{i\vec w (\vec p_1 + \vec p_2)}
\left[e^{-w_0(|p_1|+|p_2|)} +
c_{1} e^{-|p_{12}|w_0} + c_{2} e^{|p_{12}|w_0}\right]
\label{geso}
\ee
Parameter $c_2$ should vanish, $c_{2} = 0$, to prevent growth
{\it inside} the bulk. Near the boundary,  as $w_0 \rightarrow 0$,
$A_\Delta(w) \sim w_0^\Delta$. Therefore
for $\Delta =  \Delta_+ = \Delta_- + 1$ we need $c_1 = -1$
so that the asymptotics of two terms in square brackets cancel
 each other at $w_0=0$.
For $\Delta =  \Delta_-$ asymptotic itself is correct for any
$c_{1} \neq -1$, and is not enough to choose the right solution.
However, as $w_0\rightarrow 0$ the $\vec w$-Fourier transform of
$ A_{\Delta_{-}}$
should be symmetric in all the three momenta,
provided that $K_{\Delta_{-}}(w|\vec p_{1})$
and $K_{\Delta_{-}}(w| \vec p_{2})$ are chosen for the external legs:
$$
S(\vec p_1,\vec p_2,\vec p_3) \equiv
\lim_{w_0\rightarrow0} w_0^{-\Delta_-}
\int A_{\Delta_{-}}(w_0,\vec w|\vec p_1,\vec p_2)
e^{i\vec w\vec p_3}d\vec w = $$ $$ =
\frac{\delta^{(d)}(\vec p_1+\vec p_2+\vec p_3)}
{|p_1||p_2||p_3|(|p_1|+|p_2|+|p_3|)}
\times\frac{1+c_{1}}{|p_3|-|p_1|-|p_2|}
$$
$\delta^{(d)}$-function allows to substitute $|p_3| = \sqrt{\vec p_3^2}$
instead of $|p_{12}| = \sqrt{(\vec p_1+\vec p_2)^2}$.
The last ratio breaks the symmetry unless
$c_{1} = -\frac{|p_1|+|p_2|}{|p_{12}|} H(\vec p_{1}, \vec p_{2}, \vec
p_{3})$. Here $H(\vec p_{1}, \vec p_{2}, \vec p_{3})$ - is some symmetric
function of its arguments. To find this function we need to fix
the asymptotic of $A_{\Delta}(w|\vec p_{1}, \vec p_{2})\sim C w_{0}^{\Delta}$
 at $w_{0}\rightarrow 0$. This can be done by direct evaluation
of integral in
(\ref{A2++}) and in appendix C for $w_{0}=0$.
 The result is $H(\vec p_{1}, \vec p_{2}, \vec
p_{3})=1$. Thus, $c_{1} = -\frac{|p_1|+|p_2|}{|p_{12}|}$. With such choice
of $c_{1}$, for
$\Delta=\Delta_{\pm}$,  the pole at $|p_{12}| = |p_1| + |p_2|$,
i.e. at collinear momenta $\vec p_1 || \vec p_2$ disappears from the
amplitude (but only at $w_0 = 0$).

\section{Appendix C. Single-vertex diagram in {\it conventional} AdS theory
\label{phypro}}

It is instructive to repeat the calculation from s.\ref{svd} for
{\it conventional} AdS theory, with the same fields of dimensions
$\Delta_\pm$, but with the {\it physical} bulk-to-bulk propagator
\be
{\cal G}(w,z) = \left(\frac{w_0z_0}
{(w_0-z_0)^2 + (\vec w-\vec z)^2}\right)^{\Delta_-} =
(w_0z_0)^{\Delta_-}\int \frac{d^d\vec p}{|p\,|} e^{i\vec p\,(\vec w-\vec z)}
e^{-|p\,||w_0-z_0|}
\label{phpr}
\ee
instead of our projected $G_0(w,z)$ from (\ref{prod}).
The technical difference is that $w_0-z_0$ can change sign and thus
the integral over $z_0$ in the analogue of (\ref{A2++}) is more
sophisticated:
$$
{\cal A}(w|\vec p_1,\ldots,\vec p_n) \ =\
\int\frac{dz_0d^d\vec z}{z_0^{d+1}}
\ {\cal G}(w,z) K_{\Delta_+}(z|\vec p_1)\ldots K_{\Delta_+}(z|\vec p_n) \
\stackrel{(\ref{phpr}) \& (\ref{K+})}{=}\
$$ $$=\
\int\frac{dz_0d^d\vec z}{z_0^{d+1}}
\left((w_0z_0)^{\Delta_-}\int  \frac{d\vec p}{|p\,|}\
e^{i\vec p\,(\vec w-\vec z)}
e^{-|p\,||w_0-z_0|}\right)
\left(z_0^{ \Delta_-} e^{i\vec p_1\vec z} e^{-|p_1|z_0}\right)\ldots
\left(z_0^{ \Delta_-} e^{i\vec p_n\vec z} e^{-|p_n|z_0}\right)\ =
$$ $$
= \ \frac{w_0^{\Delta_-}}{|p_{1n}|}
e^{i\vec w (\vec p_1 + \ldots + \vec p_n)}
\left( e^{-|p_{1n}|w_0}\int_0^{w_0} e^{-(|p_1|+\ldots + |p_n| - |p_{1n}|)z_0}
z_0^{s_{n+1}-1}dz_0 \ + \right. $$ $$ \left. + \
e^{|p_{1n}|w_0}\int_{w_0}^{\infty} e^{-(|p_1|+\ldots + |p_n| + |p_{1n}|)z_0}
z_0^{s_{n+1}-1}dz_0\right)
\ = $$ $$ =\
\frac{w_0^{\Delta_-}}{|p_{1n}|}e^{i\vec w (\vec p_1 + \ldots + \vec p_n)}
\left\{\frac{\Gamma(s_{n+1})}{(|p_1|+\ldots + |p_n| - |p_{1n}|)^{s_{n+1}}}
e^{-|p_{1n}|w_0}\ \ +
\right. $$ $$ \left. 
\sum_{k=0}^{s_{n+1}-1}\!\!\! \frac{\Gamma(s_{n+1})}{\Gamma(s_{n+1}-k)}
w_0^{s_{n+1}-k-1}\!\left(\frac{1}{(|p_1|+\ldots + |p_n| + |p_{1n}|)^{k+1}} -
\frac{1}{(|p_1|+\ldots + |p_n| - |p_{1n}|)^{k+1}}\right)
e^{-(|p_1|+\ldots + |p_n|)w_0}\!\right\}
$$
Here $|p_{1n}| \equiv \sqrt{(\vec p_1+ \ldots + \vec p_n)^2}$.
Note that this expression has poles at $|p_{1n}| = |p_1|+\ldots + |p_n|$,
i.e. when all the $n$ momenta $\vec p_1, \ldots \vec p_n$ are collinear.

This complicated formula is typical for the amplitudes in conventional
AdS theory \cite{Rast}. In this case it is somewhat tedious, though
possible, to perform
the recursion from section \ref{mvd} and evaluate multi-vertex diagrams.
Even more difficult is to find an effective boundary theory
and the concept of ordinary AdS/CFT correspondence and/or open-closed
string duality
remains obscure, at least at the level of explicit multi-vertex
calculations.

\section{Acknowledgements}

It is a pleasure to thank
A.Alexandrov, V.Alexandrov, N.Amburg, S.Demidov, D.Gorbunov,
A.Gorsky, M.Konu\-shi\-khin, D.Levkov, M.Libanov, A.Losev, V.Lysov,
A.Mironov, M.Rotaev, V.Rubakov, G.Rubtsov, S.Sibiryakov,
 T.Tomaras and M.Vasiliev for
useful discussions.

This work is partly supported by
the Federal Program of the Russian Ministry
for Industry, Science and Technology No 40.052.1.1.1112 and the
RFBR grant 04-02-16880. The work of D.K. is also supported by the
RFBR grant 05-02-17363 and the
studentship of Dynasty Foundation.


\begin{thebibliography}{12}
\bibitem{CS} E.Witten, {\it Comm.Math.Phys.} 121,
1989, 351

\bibitem{ost}
O.Alvarez, {\it Nucl.Phys.}  B216:125, 1983; \\
S.Carlip, {\it Phys.Lett.}  B209:464, 1988; \\
A.Morozov and A.Rosly, {\it Phys.Lett.}  B195:554,
1987; {\it Phys.Lett.}  B214:533, 1988;
{\it Nucl.Phys.} B326:205, 1989;\\
S.Blau, S.Carlip, M.Clements, S.Della Pietra and
V.Della Pietra, {\it Nucl.Phys.} B301: 285, 1988;\\
A.Morozov, {\it Sov.Phys.Usp.} 35: 671-714, 1992,
({\it Usp.Fiz.Nauk} 162: 83 - 176,
http://ellib.itep.ru/mathphys/ people/morozov/92ufn-e1.ps
\& 92ufn-e2.ps

\bibitem{AdS/CFT} J.M.Maldacena,
 {\it Adv.Theor.Math.Phys.} 2:231-252,1998, Int.J.Theor.Phys.38:1113,1999
 hep-th/9711200;\\
S.Gubser, I.Klebanov and A.Polyakov, {\it Phys.Lett.} B428:105, 1998,
hep-th/9802109

\bibitem{Witt}
E.Witten, {\it Adv.Theor.Math.Phys.} 2:253-291, 1998, hep-th/9802150

\bibitem{Kl-Wit}
I.Klebanov and E.Witten,
{\it Nucl.Phys.} B556:89-114, 1999, hep-th/9905104

\bibitem{Oz}
O.Aharony, S.S.Gubser, J.M. Maldacena, H.Ooguri and Y.Oz,
{\it Phys.Rept.} 323:183, 2000, hep-th/9905111

\bibitem{Mez-Tow}
L.Mezincescu and P.K.Townsend, {\it Annals Phys.} 160:406, 1985

\bibitem{didi} A.Morozov, hep-th/9810031

\bibitem{holb} G.t'Hooft, {\it Class.Quant.Grav.} 11:621,
1994, grqc-9310006;\\
L.Susskind, {\it J.Math.Phys.}  36:6377, 1995,
hep-th/9409089

\bibitem{hole} R.Gopakumar and C.Vafa,
{\it Adv.Theor.Math.Phys.} 2:413-442, 1998, hep-th/9802016;
{\it Adv.Theor.Math.Phys.} 3:1415-1443,1999, hep-th/9811131; \\
M.Marino, hep-th/0406005; \\
A.Losev, Lectures at ITEP Schools, 2004-2005

\bibitem{Gopak}
R.Gopakumar, {\it Phys.Rev.} D70:025009, 2004,hep-th/0308184;
{\it Phys.Rev.} D70:025010, 2004, hep-th/0402063;
{\it Comptes Rendus Physique} 5:1111-1119, 2004,
hep-th/0409233; hep-th/0504229

\bibitem{amm} A.Alexandrov, A.Mironov and A.Morozov,
{\it Int.J.Mod.Phys.} A19:4127, 2004,
{\it Theor.Mat.Fiz.} 142:419, 2005,
hep-th/0310113; hep-th/0412099; hep-th/0412205;\\
A.Morozov, hep-th/0502010

\bibitem{Eyn} B.Eynard, {\it JHEP}  0411:031, 2004, hep-th/0407261; \\
L.Chekhov and B.Eynard, hep-th/0504116

\bibitem{Rub} V.Rubakov, private communications.\\
N.D.Birrell and P.C.W.Davies, "Quantum fields in Curved Space", Cambridge
University Press, 1982

\bibitem{Rast} E.D'Hoker, D.Z.Freedman and L.Rastelli,
{\it Nucl.Phys.} B562:395-411, 1999, hep-th/9905049

\bibitem{BMN}
D.Berenstein, J.M.Maldacena and H.Nastase, {\it JHEP} 0204:013, 2002,
hep-th/0202021;\\
D.Berenstein, E.Gava, J.M.Maldacena, K.S.Narain and H.Nastase,
hep-th/0203249;\\
R.R.Metsaev and A.A.Tseytlin, {\it Phys.Rev.} D65:126004, 2002,
hep-th/0202109;\\
D.J.Gross, A.Mikhailov and R.Roiban, {\it Annals Phys.} 301:31-52, 2002,
hep-th/0205066; {\it JHEP} 0305:025, 2003, hep-th/0208231;\\
N.Beisert, C.Kristjansen, J.Plefka, G.W.Semenoff, M.Staudacher,
{\it Nucl.Phys.} B650:125-161, 2003, hep-th/0208178


\bibitem{freedman}
D.Z.Freedman, S.D.Mathur, A.Matusis and L.Rastelli,
{\it Nucl.Phys.} B546:96-118, 1999, hep-th/9804058;\\
E. D'Hoker and D. Z. Freedman, {\it Nucl.Phys.} B550:261-288,1999,
hep-th/9811257;\\
E.D'Hoker, J.Erdmenger, D.Z.Freedman and M.Perez-Victoria,
{\it Nucl.Phys.} B589:3, 2000, hep-th/0003218

\bibitem{Akh} E.Akhmedov, {\it Phys.Lett.} B442:152, 1998,
hep-th/9806217; {\it Phys.Rev.} D59:101901, 1999, hep-th/9812038; \\
N.Beisert, S.Frolov, M.Staudacher, A.A.Tseytlin,
 {\it JHEP} 0310:037, 2003, hep-th/0308117;\\
A.Gorsky, {\it Theor.Math.Phys.} 142:153-165, 2005, hep-th/0308182;\\
M.Bianchi, V.Didenko, hep-th/0502220

\end{thebibliography}
\end{document}